**Evaluating Advanced Impact Mitigation Technologies for Concussion Risk Reduction in Youth Cycling Helmets**

*Jessica A. Towns[1], Fady F. Abayazid[2], Gerald A. Grant[1,3], David B. Camarillo[1], *Nicholas J. Cecchi[1]*

[1]SoftShox, Vienna, VA, USA

[2]Windsor, CA, USA

[3]Department of Neurosurgery, Duke University, Durham, NC, USA

**Corresponding author: Nicholas J. Cecchi; nick@softshox.com*

**Abstract and Key Terms**

**Purpose:** Advanced bicycle helmet technologies aim to improve concussion protection, but their efficacy under youth-specific impact conditions remains unclear. We evaluated how distinct helmet technologies affect head kinematics and predicted concussion risk and whether helmet deformation and helmet-headform relative rotation contribute to protection.

**Methods:** Six youth bicycle helmet technologies were tested, including two conventional EPS controls (FOAM-A, FOAM-B) and four advanced technologies (CELL, SLIP, SLIDE, HYDRAULIC). Helmeted headforms were impacted at three locations using a 25° anvil at 2.9 m/s, and a 45° anvil at 5.2 m/s. Peak linear acceleration (PLA), peak rotational acceleration (PRA), and YGAMBIT-predicted concussion risk were quantified. Stereo high-speed videography measured anvil-normal headform displacement to approximate helmet deformation, and helmet-headform relative rotation.

**Results:** Helmet technology significantly affected predicted concussion risk under both conditions. At 2.9 m/s, all advanced technologies reduced risk relative to FOAM-A, while only SLIDE and HYDRAULIC reduced risk relative to FOAM-B. At 5.2 m/s, all advanced technologies reduced risk relative to both foam controls. HYDRAULIC produced the lowest risk in both conditions, reducing mean risk by 94–95% and 77–78% relative to the foam controls at 2.9 and 5.2 m/s, respectively. CELL, SLIP, and SLIDE significantly reduced PRA across both conditions, whereas HYDRAULIC significantly reduced both PLA and PRA across both conditions. Greater anvil-normal headform displacement was associated with lower PLA, and greater relative rotation with lower PRA; both independently contributed to lower predicted concussion risk.

**Conclusion:** Advanced technologies in youth cycling helmets can reduce youth concussion risk, but effectiveness depends on impact condition and mitigation mechanism. Designs maximizing controlled helmet deformation, simultaneously with rotational decoupling, may provide robust protection across youth-relevant impacts.

## Introduction

Cycling is a major source of traumatic brain injury (TBI) among children and adolescents. In the United States, cycling accounts for the highest number of sports- and recreation-related TBI emergency department visits, with youth experiencing more than twice the rate of cycling-related TBI visits as adults [1,2]. Children aged 10–14 years are the most commonly affected, and male riders account for nearly three quarters of reported youth cyclist head injuries [1,2]. A recent meta-analysis of more than 48,000 youth cyclist head injuries found that concussion was the most common TBI outcome, accounting for approximately 52% of all injuries [3]. However, while helmets have been proven to significantly reduce the odds of severe head injury, concussion incidence has been found to be nearly identical between helmeted and unhelmeted adults, while helmet use has not been associated with a significant reduction in concussion odds in youth [3–8]. Improving concussion protection in youth cycling helmets therefore remains a critical opportunity for injury prevention.

Current cycling helmet test standards are not designed to evaluate concussion-relevant loading and do not incorporate youth-specific impact conditions [9]. Youth differ from adults in riding speeds and anthropometry, including body height and head mass, yet helmet certification tests use the same prescribed impact conditions for both groups. The U.S. Consumer Product Safety Commission (CPSC) standard specifies vertical drop tests in which peak linear acceleration must remain below 300 g, a criterion developed primarily to address skull fracture and severe head injuries [10,11]. This limit is considerably higher than acceleration levels associated with concussion in human studies, which report 50% injury risk values of 65–192 g [12–15]. Importantly, concussion tolerance may be even lower in youth, with Campolettano et al. reporting a mean concussive peak linear acceleration of 62 g in the only youth-specific risk analysis to our knowledge [16]. Beyond the high linear acceleration threshold, the CPSC also omits measures of rotational head kinematics, despite their established association with concussion mechanisms [17–21]. Evaluating youth helmets therefore requires a testing methodology that extends beyond the current certification criteria.

Advanced helmet technologies have increasingly been developed to mitigate both linear and rotational loading. Slip-layer systems, for instance, introduce a low-friction interface that permits relative motion between the helmet liner and head during oblique impacts, reducing transmission of tangential forces to the head and brain [22–24]. Collapsible cellular liners deform through compression, bending, and shear, allowing the same structure to accommodate normal and tangential impact forces [22,25]. Adult bicycle helmet studies have demonstrated reductions in rotational kinematics and injury metrics with slip-layer systems and collapsible cellular liners, although their effectiveness varies across helmet models, headform surface properties, and impact conditions [22,24,25]. Layered sliding-membrane systems provide another pathway for rotational decoupling by permitting controlled motion across lubricated interfaces during oblique impacts [26]. Additionally, wearable hydraulic shock absorbers dissipate impact energy by forcing fluid through restricted flow paths during compression, producing a resistive response shown to attenuate both linear and rotational forces [12,27–29]. Both sliding-membrane and hydraulic systems have reduced head kinematics during laboratory American football helmet testing, but have not yet been evaluated for cycling in a laboratory setting [12,26–28]. These findings suggest that multiple impact mitigation mechanisms may reduce head loading, but the extent to which these technologies effectively mitigate injury risk during youth-relevant cycling conditions remains unknown.

Evidence comparing advanced impact mitigation systems in youth cycling helmets remains limited. Hoshizaki et al. evaluated helmet performance across multiple impact velocities, but only one helmet was tested and the impact conditions were designed to represent tobogganing rather than cycling [30]. Jung et al. recently evaluated 21 commercially available youth helmets under oblique impacts and identified relationships between performance and design features including liner thickness, shell geometry, and presence of rotational technology [31]. However, the authors' study was designed to evaluate broader helmet design features rather than compare distinct impact attenuation mechanisms. Understanding which technologies provide the greatest protection during relevant impact conditions, and the impact attenuation mechanisms underlying their performance could better inform the design of safer youth helmets.

The present study compares four advanced youth cycling helmet technologies with conventional helmets under oblique impact conditions relevant to youth cycling crashes. We evaluated linear and rotational head kinematics, helmet deformation and relative rotation, and youth-specific predicted concussion risk. Our primary hypothesis was that the advanced technologies would reduce predicted concussion risk relative to conventional foam helmets. Our secondary hypotheses were that greater helmet deformation would be associated with lower linear head kinematics, greater helmet relative rotation would be associated with lower rotational head kinematics, and that both would be associated with reduced predicted concussion risk.

## Materials and Methods

### *Helmet Descriptions*

Six youth bicycle helmets were evaluated, representing a range of conventional and advanced linear and rotational impact mitigation technologies. Helmet sizes were selected based upon fit to the small NOCSAE headform, which has a 53.4 cm head circumference representative of a 50th-percentile ten-year-old male [32,33]. All tested helmets consisted of an exterior shell (mean thickness: 2.33 ± 0.089 mm) coupled to an expanded polystyrene (EPS) liner. Two conventional helmets featured only EPS in their liners and served as controls; one represented a lower purchase price (FOAM-A) and one represented a higher purchase price (FOAM-B). Three helmets additionally incorporated commercially available advanced impact mitigation technologies coupled to the interior of the EPS liner. One incorporated the Multi-directional Impact Protection System (i.e., MIPS), a low-friction suspended slip layer (hereafter referred to as SLIP) (Dime MIPS, GIRO; Irvine, CA); a second incorporated WaveCel, a collapsible cellular liner (CELL) (Jet Wavecel, Bontrager; Waterloo, WI); and a third incorporated the DVRT sliding membrane liner (SLIDE) (Hendrix Jr. DVRT, Bern Helmets; Plymouth, MA). A fourth helmet incorporating an advanced impact mitigation technology consisted of a prototype hydraulic liner with SoftShox wearable hydraulic shock absorbers (HYDRAULIC) (SoftShox; Vienna, VA). The HYDRAULIC prototype was created by removing 10 mm of EPS from the space between the FOAM-A liner and helmet shell, and replacing this space with nine wearable hydraulic shock absorbers while preserving the original total liner thickness. Together, these configurations were selected to represent distinct mechanisms for managing impact energy within the limited range of advanced technologies currently available in youth cycling helmets. Helmet cross-sections are depicted in Figure 1. Retail prices for commercially available helmets ranged from $29.99 to $89.99 USD. Further design characteristics are provided in Table 1.

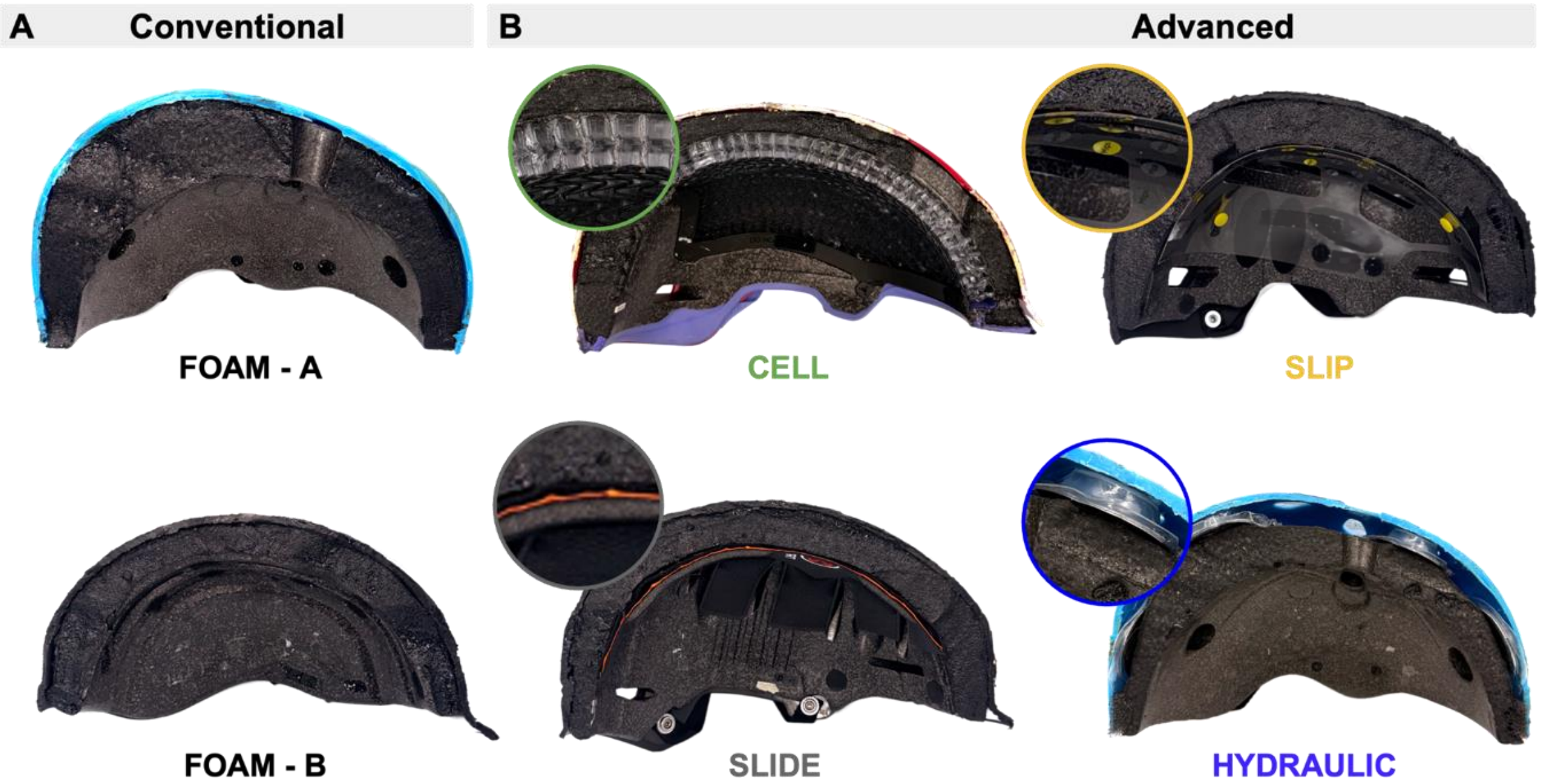

**Fig. 1** Cross-sectional views of the youth cycling helmets tested in this study, with magnified detail views highlighting differences among the advanced impact mitigation liner components. A) Conventional expanded polystyrene (EPS) foam helmets at a lower price point (FOAM-A) and a higher price point (FOAM-B). B) Advanced helmet technologies, including a collapsible cellular liner (CELL), suspended slip liner (SLIP), sliding membrane liner (SLIDE), and a prototype liner with wearable hydraulic shock absorbers (HYDRAULIC)

### *Impact Setup*

Impact tests were conducted using a monorail drop tower (CADEX, Quebec, Canada) equipped with a custom halo fixture featuring six adjustable prongs for positioning the headform (Figure 2A). A small NOCSAE headform was selected to represent the target age group because males approximately 10-14 years old comprise the most prevalent demographic for pediatric cycling-related head injuries [1,3]. Three impact locations were evaluated, including the front boss, crown, and rear boss. Impact locations were adapted from the Virginia Tech Bicycle STAR helmet testing protocol [34] (Figure 2).

| Helmet ID | Model | Weight (g) | Shell Thickness (mm) | Size | Retail Price (USD) |
|---|---|---|---|---|---|
| FOAM - A | Nattork Classic | 389 | 2.35 | Medium | $29.99 |
| FOAM - B | Thousand Next | 397 | 2.20 | Medium | $84.95 |
| CELL | Bontrager Jet WaveCel | 559 | 2.39 | Youth | $89.99 |
| SLIP | GIRO Dime MIPS | 411 | 2.24 | Small | $74.95 |
| SLIDE | Bern Hendrix Jr. DVRT | 407 | 2.43 | Small | $79.00 |
| HYDRAULIC | SoftShox Prototype | 488 | 2.35 | Medium | N/A |

**Table 1.** Helmet weights and shell thicknesses were measured prior to impact testing. Additional details were gathered from the product websites at time of purchase.

Each helmet was positioned on the headform using the small NOCSAE nose gauge. To accurately position the helmeted headform for each location, headform orientation was prescribed independently for each impact location and reproduced before every test. The axes of the positioning coordinate system followed SAE J211 conventions from the perspective of the operator facing the monorail [35]. Rotation about the Z axis was measured using angular markings inscribed in 5° increments on the positioning halo, with 0° defined along the radial direction from the center of the halo to the anvil apex. Rotations about the X and Y axes were measured using a dual-axis inclinometer placed in the headform neck opening. All impacts were conducted within ±0.5° of the prescribed orientation in each axis (Table 2).

| Impact Location | X (deg) | Y (deg) | Z (deg) |
| --- | --- | --- | --- |
| Front boss | 17.2 | 1.7 | -75 |
| Crown | -7.0 | 43.7 | 15 |
| Rear boss | 2.6 | 12.2 | -110 |

**Table 2**. Orientations corresponding to each impact location. X and Y were determined by a dual-axis inclinometer positioned in the neck opening of the NOCSAE headform. Z was determined by aligning the sagittal line on the face of the NOCSAE headform with the halo markings, inscribed in 5° increments. Impact locations are adapted from the Virginia Tech Bicycle STAR protocol [34].

Two impact conditions were evaluated, representing different impact severities. These included a 2.9 m/s drop against a 25° angled anvil and 5.2 m/s drop against a 45° angled anvil. The anvil surface was covered with 80-grit sandpaper to represent the road surface [34,36]. Prior work has shown that cycling head impacts frequently occur at angles between approximately 20° and 60°, and that impact angle increases with speed, motivating selection of the two velocity-anvil conditions [37–40]. The maximum theoretical head impact velocity for a ten-year-old rider was estimated from the gravitational potential energy during a fall:

$$V_{max} = \sqrt{2gh} \quad (1)$$

where (g) is the acceleration due to gravity (9.81 m/s$^2$) and (h) is seated head height above ground. For a 50th percentile ten-year-old male (1.404 m standing height, 0.7067 m inseam) seated head height was calculated as: upper body height (standing height minus inseam) plus saddle height above ground. Saddle height above ground was calculated as bottom bracket height plus 0.875 x inseam, based on children's preferred cycling posture and consistent with Lemond-Guimard method [41–44]. Standing height corresponds to a 24-inch bicycle, for which bottom bracket height was assumed to be 0.3 m based on commercially available 24-inch bicycles (general-use, train, and mountain bicycle designs) [42]. Maximum impact velocity was therefore 5.64 m/s, with the 2.9 and 5.2 m/s test conditions representing 50% and 90% of this velocity, respectively.

Each helmet specimen was assigned to only one velocity-anvil condition and was impacted once at each of the three locations. Helmet configurations were tested in the order FOAM-A, CELL, SLIP, SLIDE, HYDRAULIC, and FOAM-B, and locations were tested in the order front boss, crown, and rear boss. Two replicate helmet specimens were initially tested for each configuration at each severity condition, resulting in 72 planned impacts across 24 helmets, with four helmet specimens per configuration. Two replicates were selected based on an *a priori* power analysis providing 80% power at α=0.05 to detect a 10-percentage-point difference in predicted concussion risk between technologies. Prior studies testing cycling helmets have similarly included two replicates [24,31,34].

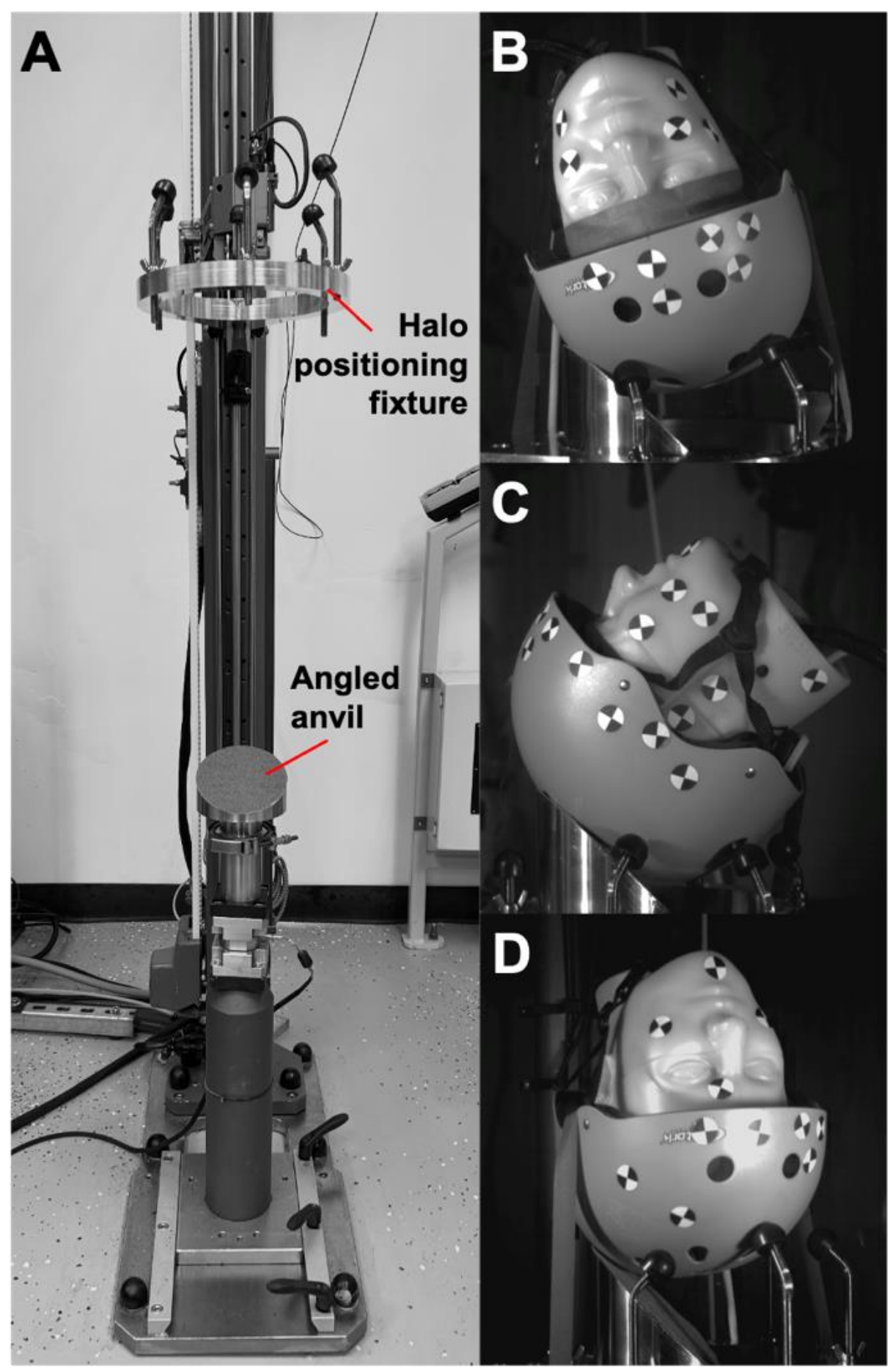

**Fig. 2** Oblique helmet testing setup. **A**) Monorail system used to test **B)** front boss, **C)** crown, and **D)** rear boss impact locations

A 6DX-PRO sensor (Diversified Technical Systems, Seal Beach, CA) was mounted at the center of gravity of the NOCSAE headform to measure triaxial linear acceleration and triaxial rotational velocity during each impact. Linear kinematics were filtered at CFC1000 according to SAE J211 specifications. Rotational acceleration was obtained from the rotational velocity signal using a five-point stencil derivative, with both filtered at CFC180. Peak linear acceleration (PLA) and peak rotational acceleration (PRA) were used to calculate YGAMBIT, a concussion risk function derived from youth football concussions [16]. YGAMBIT values were transformed to concussion risk values with a lognormal cumulative distribution with parameters μ = 0.967 and $\sigma$ = 0.331:

$$YGAMBIT = \sqrt{\left(\frac{PLA}{62.4}\right)^2 + \left(\frac{PRA}{2609}\right)^2} \quad (\mathbf{2})$$

$$Concussion\ risk = \frac{1}{2} + \frac{1}{2} erf\left[\frac{\ln(YGAMBIT) - 0.967}{\sqrt{2} \times 0.331}\right] \quad (3)$$

### *High Speed Video Measurement*

High-speed stereo video was used to quantify helmet-headform relative rotation and headform displacement toward the anvil during impact. Three-dimensional motion of the helmet and headform was reconstructed using the methods described below.

Impacts were recorded at 3,600 frames per second using two high-speed video cameras (FASTCAM SA3, Photron, Tokyo, Japan) positioned to capture the primary axis of head rotation, while maintaining overlapping fields of view for stereo 3D reconstruction of the helmet and headform. Multiple 20 mm tracking markers were placed on both the helmet and headform to be used as tracking points during analysis (Figure 3A).

The stereo camera system was calibrated before testing using images of a planar checkerboard moved throughout the reconstruction volume. Intrinsic and extrinsic camera parameters were estimated using the MATLAB Stereo Camera Calibrator application (MathWorks; Natick, MA). Marker trajectories were manually tracked in both camera views using Tracker Video Analysis and Modelling Tool (Tracker 4.91; Open Source Physics) and triangulated in MATLAB to obtain three-dimensional marker coordinates [45]. Reconstructed marker positions were reprojected into each camera image as a quality-control measure, and markers with reprojection errors exceeding 5 pixels were excluded.

The helmet and headform were each modeled as rigid bodies. At least three non-collinear markers were required to define each rigid-body orientation. For each video frame, the best-fit rotation between the reference and current marker configurations was determined by singular value decomposition. Marker clusters were selected such that maximum inter-marker distance changes remained below 2 mm, avoiding regions susceptible to substantial local shell deformation, including the immediate impact site and shell brim.

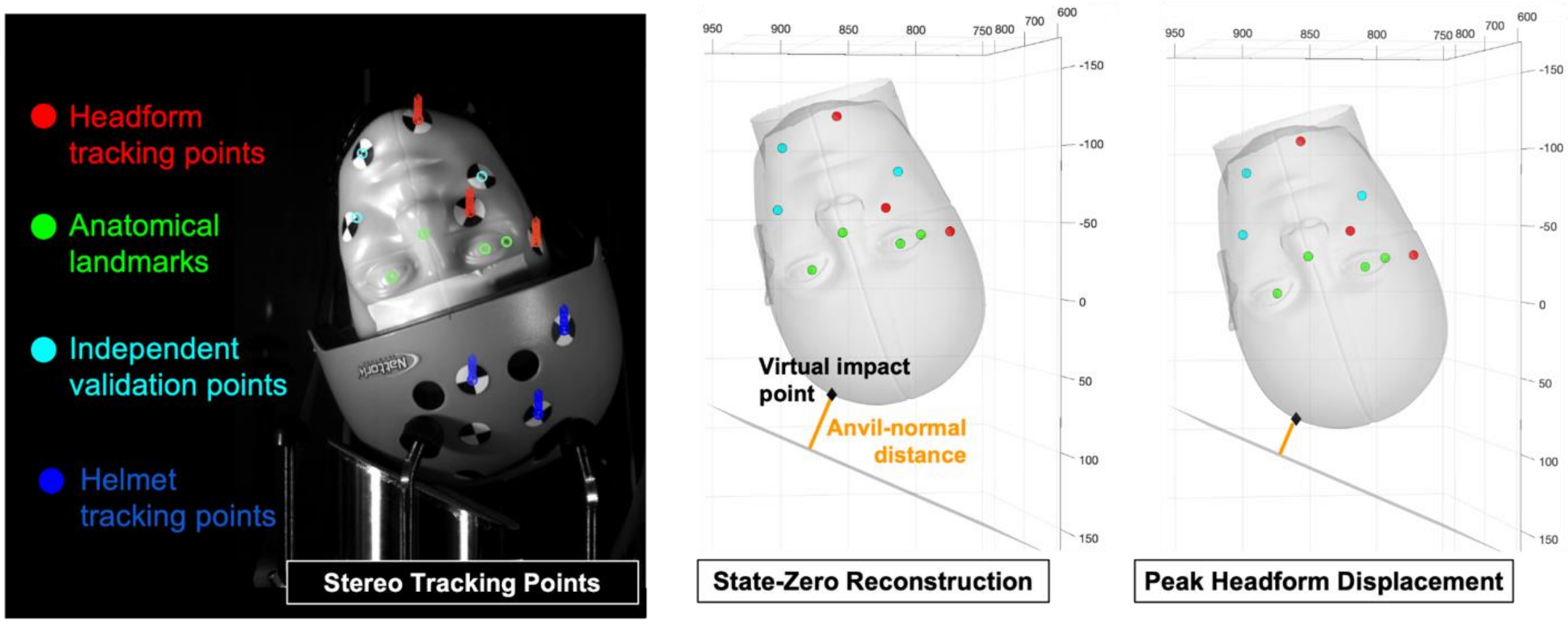

**Fig. 3** Stereo high-speed video reconstruction of headform motion. A) Representative stereo tracking configuration for a front boss impact to a 25° anvil. B) State-zero reconstruction after registration of the NOCSAE headform mesh. C) Headform pose at peak anvil-normal displacement, obtained by propagating rigid-body headform motion from state zero

A separate state-zero reconstruction was performed for each impact to establish the initial three-dimensional position and orientation of the headform relative to the anvil (Figure 3A-B). Within the 5 ms immediately preceding first contact with the anvil, the video frame producing the lowest mean reprojection error across the tracked headform markers was selected. At this frame, anatomical landmarks visible in both camera views and identifiable on a meshed 3D scan (Einstar, Shining 3D; Hangzhou, China) of the NOCSAE headform were manually selected and reconstructed. At least three non-collinear anatomical landmarks were used to rigidly register the headform mesh to its reconstructed video position using a least-squares singular-value-decomposition solution.

Registration quality was quantified using the RMSE and maximum residual distances between the reconstructed anatomical landmarks and their corresponding locations on the registered mesh, as well as differences in pairwise landmark distances. Additional reconstructed headform surface points that were not used to perform the registration served as independent checks of the reconstructed STL position.

The anvil plane and surface-normal direction were then reconstructed within the stereo coordinate system using the known laboratory geometry and anvil angle. Because the headform remained above the anvil at state zero and was separated from the anvil by the helmet, it was translated vertically downward until first contact with the reconstructed anvil plane. The corresponding surface point was defined as the virtual impact point (Figure 3B). This point was fixed relative to the headform rigid body marker cluster at state zero and propagated throughout the impact using the stereo-measured motion of the headform. This approach allowed motion at the local helmet-headform interface to be quantified despite its occlusion during impact.

Video-derived headform rotational velocity was calculated from successive orientation matrices and low-pass filtered at 235 Hz. The video rotational velocity baseline was corrected by subtracting the mean of four samples preceding impact onset. Because video and sensor recordings did not share a trigger, time series data were synchronized by cross-correlating the video- and sensor-measured headform rotational velocity traces. The lag corresponding to maximum cross-correlation was applied to the video time axis, while the sensor time axis was retained as the reference. The accuracy of the stereo rigid-body reconstruction was evaluated by comparing video- and sensor-measured peak resultant rotational velocity using root mean square error (RMSE) and mean absolute error (MAE).

For helmet-headform dynamics calculations, the reference position was defined at the first rising crossing of 15 g in the sensor-measured resultant linear-acceleration trace. This objectively defined reference matches the accelerometer trigger threshold, and was chosen to reduce the contribution of early displacement associated with differences in helmet fit or initial air gaps between the headform and liner, which could affect a subjective, visual contact-based definition of impact onset. Helmet-headform relative rotation was calculated from the change in helmet orientation relative to the headform reference frame and expressed as a three-dimensional rotation vector. The magnitude of this vector was used to determine maximum helmet-headform relative rotation.

Headform displacement toward the anvil-normal was calculated by propagating the virtual impact point with the measured headform motion and projecting its displacement onto the anvil surface normal. Displacement and relative-rotation signals were low-pass filtered at 235 Hz, and peak anvil-normal displacement and peak helmet-headform relative rotation were extracted.

***Statistical Analysis***

Two replicate impacts were initially collected for each technology across all impact conditions. For replicate quality control, the absolute difference in YGAMBIT-predicted concussion risk between the two replicate measurements was calculated for each technology. This was done per velocity-anvil condition, producing a distribution of replicate differences separately for the 2.9 m/s, 25° and 5.2 m/s, 45° conditions. Normality of each distribution was assessed using the Shapiro-Wilk test. Because the replicate-difference distributions were non-normal, potential outliers were identified using Tukey's 1.5×IQR criterion, with values below Q1 − 1.5×IQR or above Q3 + 1.5×IQR classified as outliers. When a replicate pair was flagged as an outlier, a third impact was conducted and the two replicates with the smallest difference in predicted concussion risk were retained for statistical analysis. This approach identified four outlier pairs, for which a third impact was conducted and the two closest replicates were retained, yielding 72 impacts across the 6 helmet technologies. Prior studies have similarly used outlier detection in impact testing to identify and replace discordant trials [46,47].

Differences in YGAMBIT-predicted concussion risk, PLA, and PRA among helmet technologies were evaluated separately for the two velocity-anvil conditions using two-way analyses of variance (ANOVAs), with helmet technology and impact location as factors (GraphPad Prism 11.1.0, Boston, MA). Tukey's multiple-comparisons tests compared technology means averaged across impact locations. Statistical significance was defined as $\alpha = 0.05$. Post hoc comparisons for risk differences were performed following a significant technology-by-location interaction to compare helmet technologies within each impact location. Pairwise comparisons were again conducted separately for each velocity-anvil condition using a two-way ANOVA with Tukey's multiple comparisons test.

Shapley decomposition was used to determine the contributions of PLA and PRA reductions to the reduction in predicted concussion risk relative to each FOAM control. Shapley contributions were normalized to the mean predicted risk of the corresponding FOAM control and expressed as percent reductions. Additionally, percent reductions in headform kinematics were calculated from technology-specific means across impact locations within each condition.

The relationship between helmet-headform motion and peak kinematics was evaluated using two linear regression models (statsmodels 0.14.6). The first model included peak headform anvil-normal displacement as a predictor of PLA (Equation 4), while the second model included maximum helmet-headform relative rotation as a predictor of PRA (Equation 5). Both models adjusted for impact location and a categorical "impact condition" variable representing the paired impact velocity and anvil angle (45° and 5.2 m/s; 25° and 2.9 m/s) experiments:

$$PLA = \beta_0 + \beta_1(Peak\ Displacement) + \beta_2(Impact\ Condition) + \beta_3(Impact\ Location) \quad (\mathbf{4})$$

$$PRA = \beta_0 + \beta_1(Peak\ Relative\ Rotation) + \beta_2(Impact\ Condition) + \beta_3(Impact\ Location) \quad (\mathbf{5})$$

The relationship between helmet-headform motion and predicted concussion risk was evaluated using a beta regression with a logit link, given risk is a continuous outcome bounded between 0 and 1. Both measures were included simultaneously, with adjustment for impact condition and impact location:

$$logit(E[Concussion\ Risk]) = \beta_0 + \beta_1(Peak\ Displacement) + \beta_2(Peak\ Relative\ Rotation) + \beta_2(Impact\ Condition) + \beta_3(Impact\ Location) \quad (\mathbf{6})$$

The independent contribution of displacement and relative rotation to concussion risk was evaluated using Wald z-tests and likelihood-ratio tests comparing the combined model with nested models omitting each stereo-measured predictor. Helmet technology was not included as a covariate because the objective was to characterize the underlying relationship between measured helmet mechanics and predicted concussion risk. Beta regression models were fit by maximum likelihood, and linear regression models were fit by ordinary least squares.

## Results

### *Advanced technologies significantly reduce predicted concussion risk.*

Across both impact conditions, predicted concussion risk varied significantly by helmet technology and impact location, with a significant interaction between the two factors. Figure 4 shows results for the 2.9 m/s, 25° anvil condition, while Figure 5 shows results for the 5.2 m/s, 45° anvil condition. Both figures present the mean and standard deviation of concussion risk for each helmet technology at each impact location, along with the mean peak kinematics associated with each technology's mean concussion risk. Summary statistics for each technology are provided in Supplementary Table 1.

At 2.9 m/s and 25°, concussion risk differed significantly among helmet technologies, $F(5,18) = 127.9$, $p<0.0001$, and impact locations, $F(2,18) = 143.6$, $p<0.0001$, with a significant technology-by-location interaction, $F(10,18) = 33.41$, $p<0.0001$. All helmet technologies had significantly lower mean concussion risk than FOAM-A. Relative to FOAM-B, only SLIDE and HYDRAULIC produced significant reductions, whereas CELL and SLIP did not differ significantly from FOAM-B. Among the advanced technologies, HYDRAULIC produced the lowest mean concussion risk, with reductions of 94.1–95.3% relative to FOAM and and 90.9%–93.9% relative to CELL, SLIP, and SLIDE, and was significantly lower than each of these five technologies (all $p<0.0001$). Pairwise comparisons are provided in Supplementary Table 2.

At 5.2 m/s and 45°, concussion risk again differed significantly among helmet technologies, $F(5,18) = 269.2$, $p<0.0001$, and impact locations, $F(2,18) = 165.6$, $p<0.0001$, with a significant technology-by-location interaction, $F(10,18) = 11.47$, $p<0.0001$. FOAM-A and FOAM-B did not differ significantly from one another, while CELL, SLIP, SLIDE, and HYDRAULIC each produced significantly lower mean concussion risk than both FOAM-A and FOAM-B. Among the advanced technologies, HYDRAULIC produced the lowest mean concussion risk and was significantly lower than all tested technologies (all $p<0.0001$), with reductions of 76.9–77.5% relative to FOAM-A, and FOAM-B and 57.1%–67.9% relative to CELL, SLIP, and SLIDE. Pairwise comparisons are provided in Supplementary Table 3.

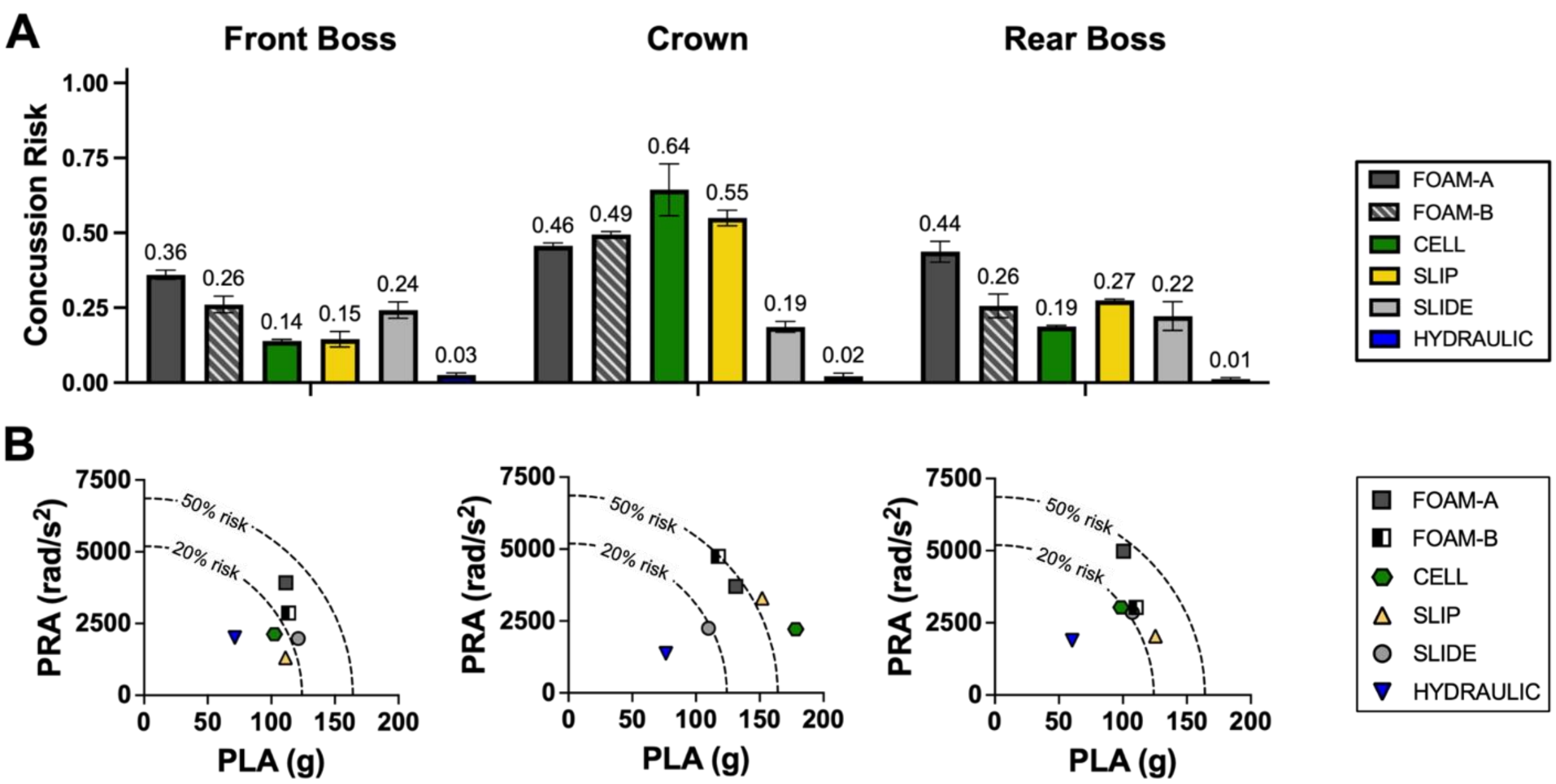

**Fig. 4** Impacts conducted at 25° and 2.9 m/s. A) Predicted concussion risk for each helmet technology at the front boss, crown, and rear boss impact locations. Bars represent the mean of two replicates, and error bars denote the standard deviation. B) Mean peak linear acceleration and peak rotational acceleration associated with mean concussion risk depicted above. The 20% and 50% risk contours are overlaid as dashed lines for reference

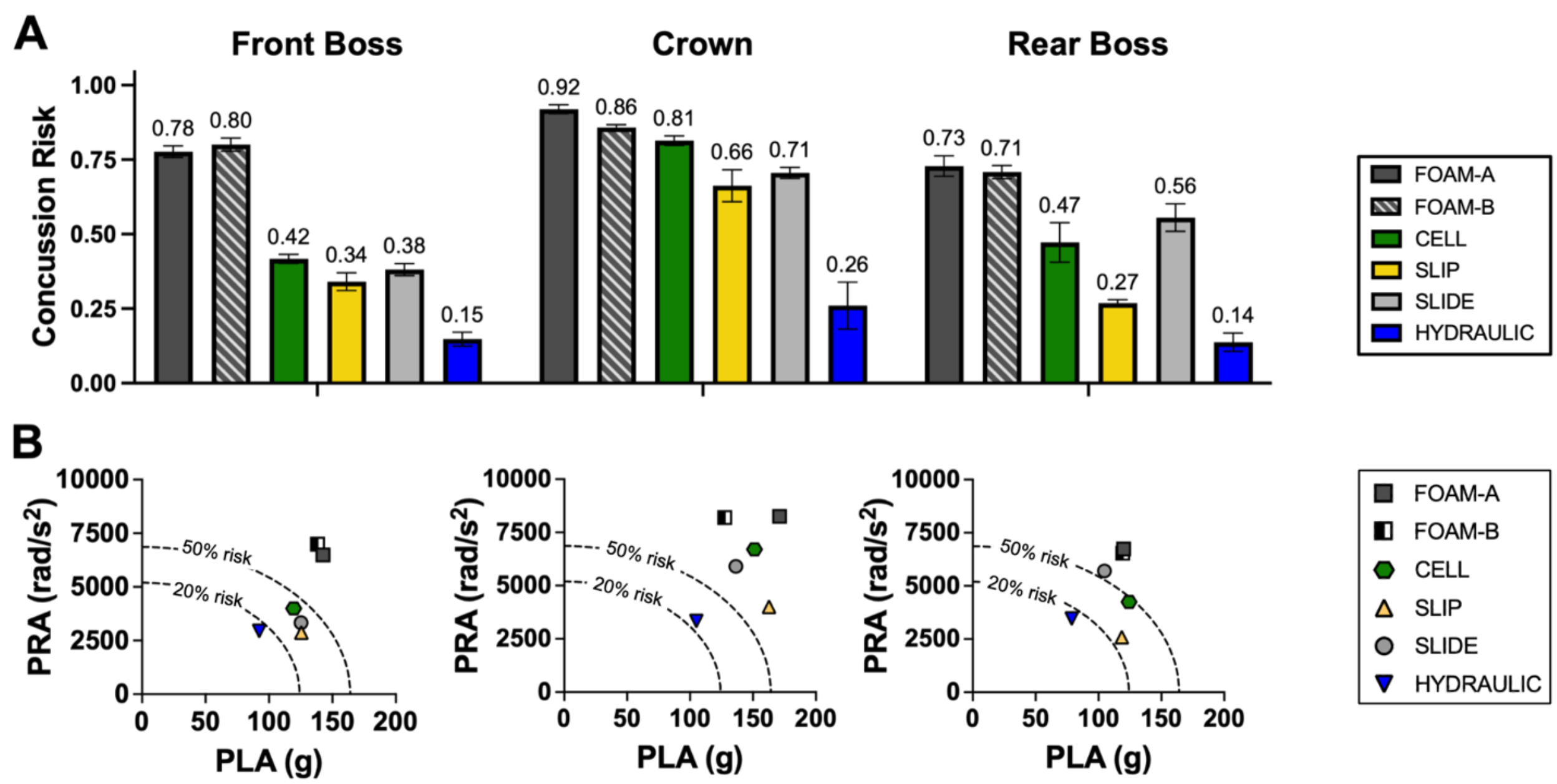

**Fig. 5** Impacts conducted at 45° and 5.2 m/s. A) Predicted concussion risk for each helmet technology at the front boss, crown, and rear boss impact locations. Bars represent the mean of two replicates, and error bars denote the standard deviation. B) Mean peak linear acceleration and peak rotational acceleration associated with mean concussion risk depicted above. The 20% and 50% risk contours are overlaid as dashed lines for reference

The significant technology-by-location interaction was reflected in location-dependent differences between helmet technologies. At 2.9 m/s, CELL produced significantly higher predicted concussion risk than both FOAM helmets at the crown but significantly lower risk at the front boss and rear boss. SLIP significantly reduced risk relative to both FOAM helmets at the front boss only, whereas SLIDE produced significant reductions at the crown and rear boss. At 5.2 m/s, CELL significantly reduced risk relative to both FOAM helmets at the front boss and rear boss, SLIP reduced risk at the front boss and rear boss, and SLIDE reduced risk at the crown and rear boss. In contrast, HYDRAULIC significantly reduced predicted concussion risk relative to both FOAM helmets at all three impact locations under both conditions (all $p<0.0001$). Pairwise comparisons for 2.9 and 5.2 m/s conditions are provided in Supplementary Tables 4 and 5, respectively.

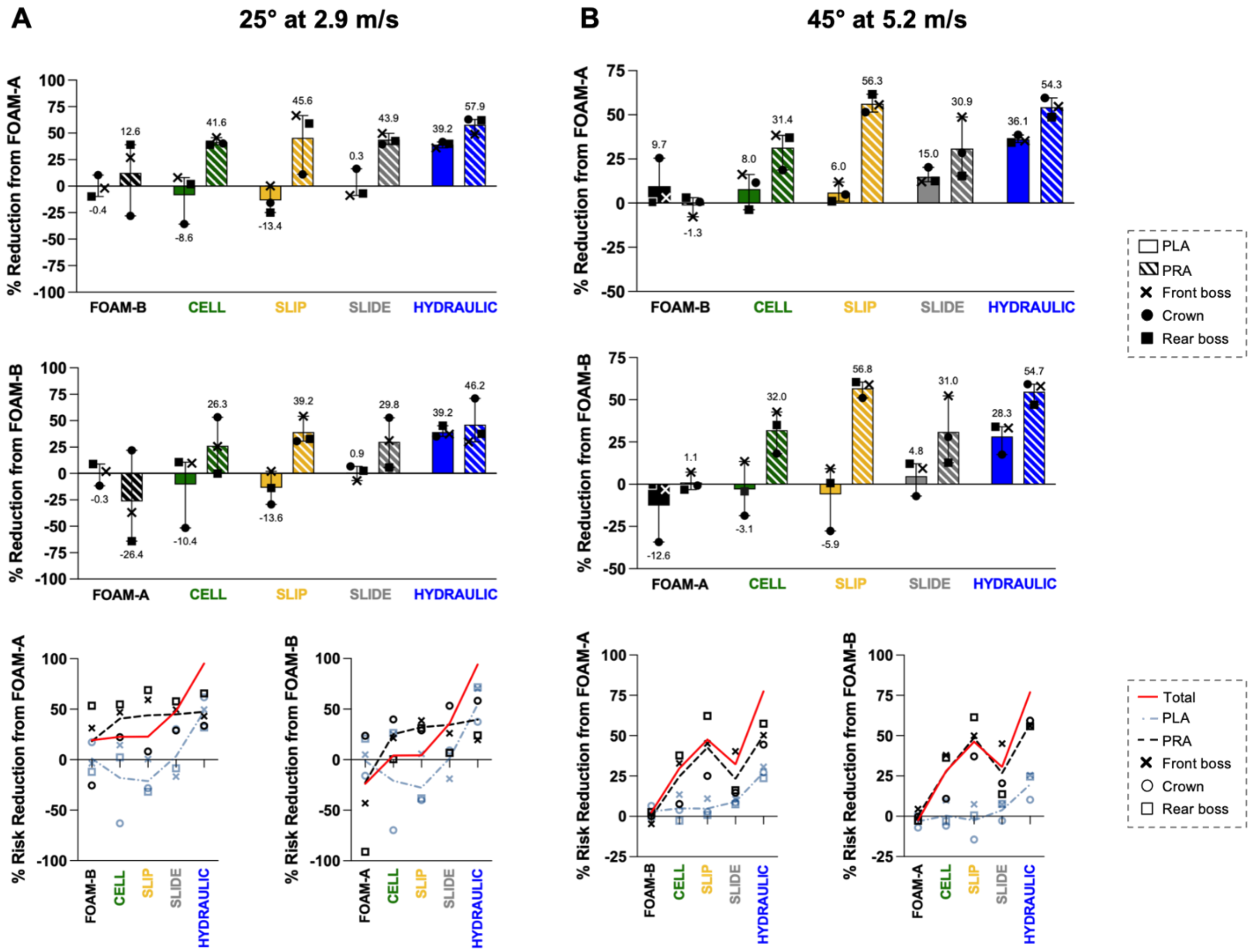

**Fig. 6** Percent reductions in PLA, PRA, and predicted concussion risk relative to FOAM-A and FOAM-B controls at A) 25° and 2.9 m/s, and B) 45° and 5.2 m/s. Top and middle rows show reductions in PLA and PRA, with bars showing means calculated across all observations within each velocity-anvil condition. The bottom row shows the relative reduction in predicted concussion risk, broken down by reductions due to PLA and PRA determined using Shapley decomposition, with lines showing means calculated across all observations within each velocity-anvil condition. Symbols in all rows show means calculated separately for each impact location

### *Attenuation of linear acceleration can further improve concussion risk reductions.*

PLA and PRA differed significantly across helmet technologies for both velocity-anvil conditions, with ANOVA results reported in Supplementary Tables 6-9. Kinematic reductions from FOAM differed across helmet technologies and were generally greater for rotational than linear acceleration (Figure 6). All technologies significantly reduced PRA from FOAM for both conditions. At 25°, CELL and SLIP reduced PRA by 29.5–47.1% despite increasing PLA by 13.5–15.4%. SLIDE reduced PLA by -0.80–0.50% and PRA by 30.1–43.0%. At 45°, CELL and SLIP reduced PLA by -1.4–8.9% and PRA by 30.7–56.2%. SLIDE reduced PLA by 8.1–16.2% and PRA by 31% at 45°. HYDRAULIC was the only technology to significantly reduce PLA and PRA across both conditions, lowering PLA by 30.7–37.8% and PRA by 48.1–57.7%.

Kinematic summaries across technologies and impact conditions are provided in Supplementary Table 1. At 2.9 m/s, mean PLA was 111.1 ± 27.1 g, while mean PRA was 2756 ± 1064 rad/s$^2$. At 5.2 m/s, mean PLA was 125.7 ± 23.1 g, while mean PRA was 5128 ± 1888 rad/s$^2$.

***Helmet-headform relative motion was associated with lower kinematics and concussion risk.***

Stereo-motion analyses yielded high reconstruction quality, with a mean reprojection error of 0.96 ± 0.53 pixels and mean maximum rigid-body residuals of 0.26 ± 0.27 mm for the helmet and 0.30 ± 0.10 mm for the headform. Video-derived PRV showed close agreement with the sensor measurements, with an RMSE of 3.18 rad/s and MAE of 2.31 rad/s.

At 25° and 2.9 m/s, peak anvil-normal displacement averaged 12.68 ± 3.47 mm and peak relative rotation averaged 5.88 ± 3.01°. At 45° and 5.2 m/s, these averaged 17.14 ± 2.94 mm and 12.52 ± 5.22°, respectively. Figure 7 shows representative snapshots from high-speed video collected during front boss impacts to the 25° anvil, where HYDRAULIC exhibited visibly greater helmet-headform motion by 15 ms.

Stereo-measured helmet-headform relative motion was significantly associated with headform kinematics Figure 8). Greater headform displacement toward the anvil was associated with lower PLA ($\beta = -4.97$ g/mm, 95% CI [−7.21, −2.73], $p = 9.77 \times 10^{-5}$). Greater helmet-headform relative rotation was associated with lower PRA ($\beta = -288.5$ rad/s$^2$/°, 95% CI [−422.7, −154.4], $p = 1.41 \times 10^{-4}$).

Greater helmet-headform displacement and relative rotation were each associated with lower predicted concussion risk. In the beta-regression model including both measures, greater displacement was associated with lower predicted concussion risk ($\beta = -0.192$ per mm, 95% CI [−0.277, −0.106], $p = 1.02 \times 10^{-5}$), as was greater relative rotation ($\beta = -0.110$ per degree, 95% CI [−0.193, −0.027], $p = 0.0097$). Together, the two motion measures significantly improved prediction beyond impact condition and location ($\chi^2 = 33.21$, $p = 6.16 \times 10^{-8}$). When evaluating the independent contribution of each motion predictor, adding relative rotation to a model containing displacement improved fit ($\chi^2 = 6.31$, $p = 0.012$), while adding displacement to the relative rotation model also improved fit ($\chi^2 = 16.79$, $p = 4.19 \times 10^{-5}$), establishing that each provided independent information about concussion risk.

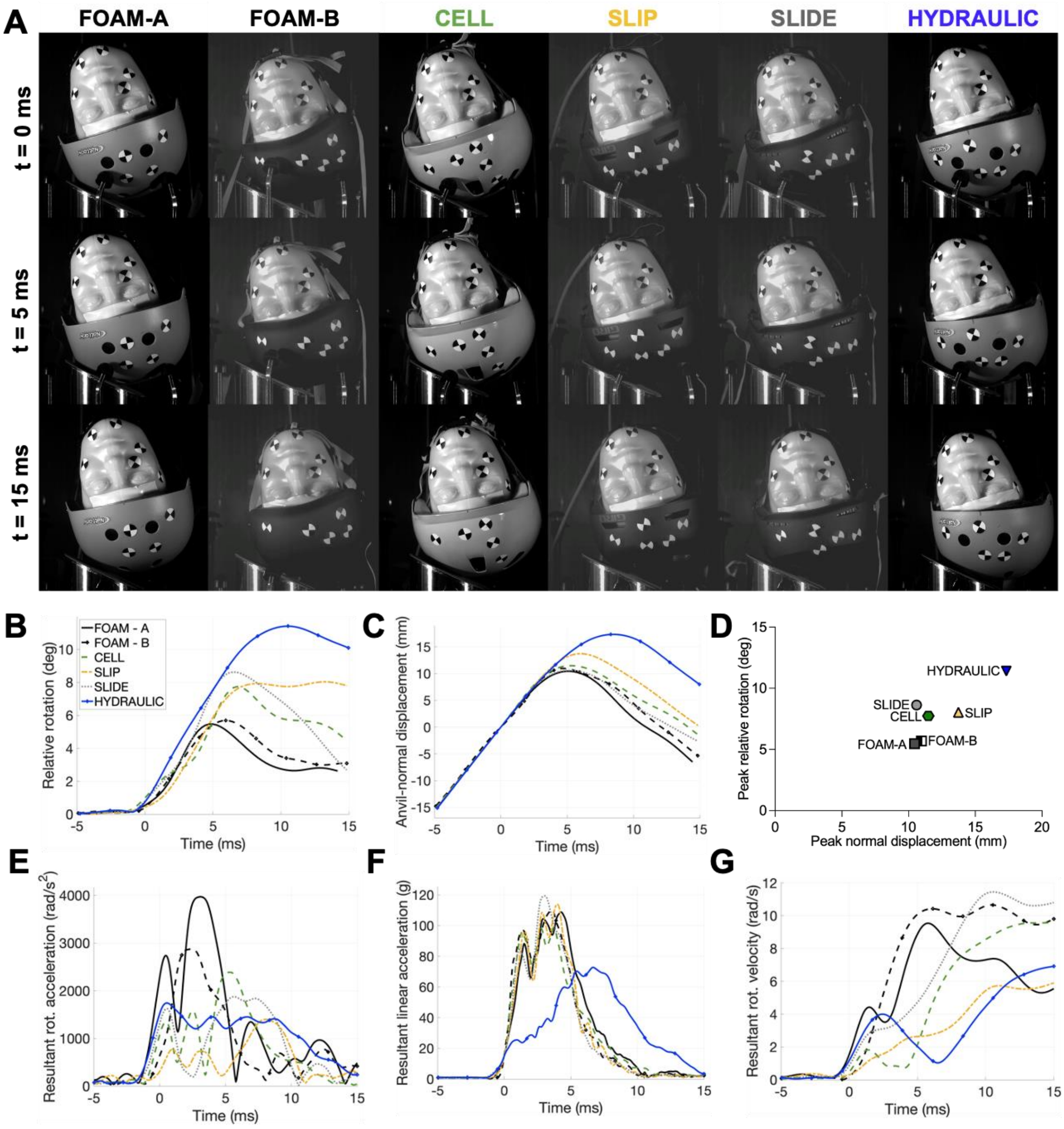

**Fig. 7** Helmet-headform relative motion during front boss impacts to the 25° anvil at 2.9 m/s. A) Representative high-speed video frames at 0, 5, and 15 ms illustrate differences in helmet motion among technologies. B) Resultant helmet-headform relative rotation, C) anvil-normal headform displacement, and D) their corresponding peak values. E–G) Corresponding sensor-measured headform rotational velocity, linear acceleration, and rotational acceleration, respectively

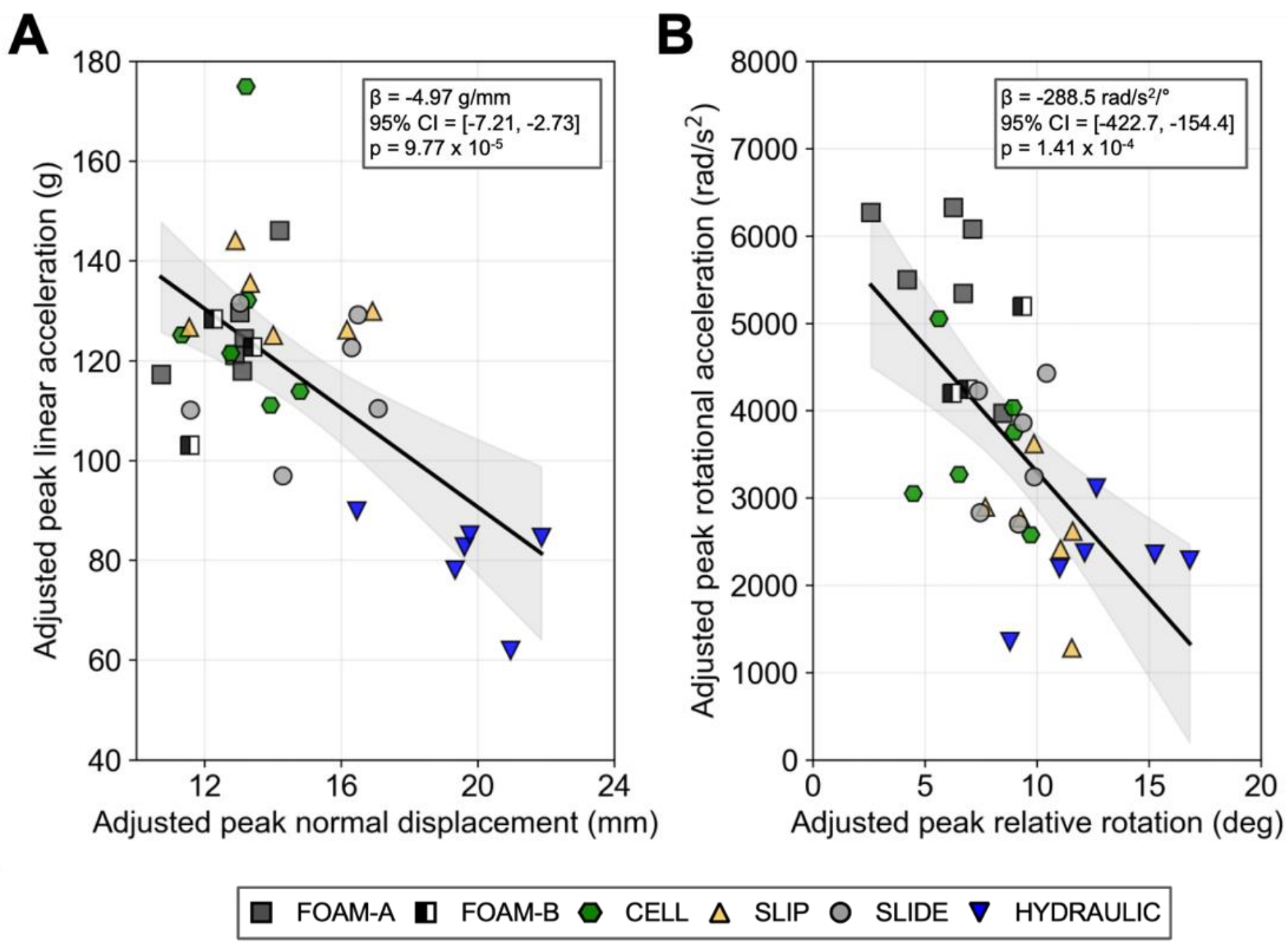

**Fig. 8** Linear regression models relating stereo-measured helmet-headform motion to headform kinematics. A) Association between peak normal displacement and peak linear acceleration, and B) peak relative rotation and peak rotational acceleration. Plotted points were adjusted for impact condition and impact location, and do not represent raw observations. Lines show the adjusted regression relationships, with shaded regions indicating 95% confidence intervals

## Discussion

To our knowledge, this is the first study to systematically compare advanced bicycle helmet technologies under youth-specific impact conditions, and the first to quantify an association between helmet-headform dynamics and concussion risk. Conventional EPS foam helmets produced concussion risk as high as 92%, but this was significantly reduced by all advanced technologies. A helmet featuring HYDRAULIC technology produced the greatest reduction, lowering concussion risk by up 95% relative to FOAM through attenuation of both linear and rotational acceleration. Based on impact stereo reconstruction, greater headform displacement toward the anvil was associated with lower linear head acceleration, while greater helmet-headform relative rotation was associated with lower rotational head acceleration. Both motions independently contributed to lower predicted concussion risk, confirming complementary mechanisms through which helmet technologies can reduce concussion-relevant head loading.

Predicted concussion risk remained substantial in conventional foam helmets, reaching 26–50% in the lower-severity impact condition and 70–91% in the higher-severity impact condition. FOAM-B outperformed FOAM-A only at 2.9 m/s, and both performed similarly at 5.2 m/s. This likely reflects differences in EPS density and thickness, and shell properties, considering that foam responses may be tuned for specific impact energies [48]. Nevertheless, every advanced technology significantly lowered predicted risk relative to FOAM-B for at least one impact condition, despite being offered at similar price points, and all significantly outperformed FOAM-A during both impact conditions. The HYDRAULIC helmet produced the largest and most consistent benefit, reducing predicted concussion risk by 94–95% relative to both FOAM helmets at 2.9 m/s and by 77–78% at 5.2 m/s. These findings suggest that advanced impact mitigation technologies can provide meaningful improvements in concussion protection beyond those achieved by conventional EPS alone. Notably, FOAM-B's limited performance advantage occurred despite its higher price point relative to FOAM-A, and advanced technologies achieved superior performance at similar price points to FOAM-B. Prior bicycle helmet studies have likewise found that price is a weak or inconsistent indicator of protective performance, with considerable performance variation observed among helmets at similar price points [34,49,50]. Our results suggest that performance improvements may be achievable without requiring consumers to move into a higher price tier, although this relationship should be evaluated across a larger number of helmet models.

The significant technology-by-location interaction and post-hoc comparisons showed that protection depended on where the helmet was impacted for most technologies. For example, at 2.9 m/s, CELL increased risk at the crown location relative to both FOAM helmets, while reducing risk at the remaining two locations. For this same condition, SLIP only provided significant risk reductions from both FOAM helmets at the front boss location, and SLIDE at the crown and rear boss locations. In contrast, HYDRAULIC provided a consistent reduction from FOAM across all impact locations and conditions. These location-dependent differences may arise because local shell curvature changes the direction in which impact loading is transferred through the helmet, which could affect how effectively different mitigation mechanisms respond. Local differences in the underlying EPS thickness could also alter this response. HYDRAULIC may be less sensitive to these local effects because the shell distributes loading across the discrete shock absorber units positioned between the shell and EPS liner. For CELL, anisotropy of its cellular liner could further drive location-dependent differences in deformation, as similar effects have been reported for MIPS and WaveCel during oblique bicycle helmet impacts [22,24]. Future work should examine how local helmet geometry and loading direction interact with specific impact mitigation mechanisms to identify design features that preserve protection across a broader range of locations.

Kinematic differences between technologies show that risk reductions were achieved through different balances of linear and rotational attenuation. CELL, SLIDE, and SLIP technologies favored rotational attenuation, reducing PRA more than PLA across conditions. At 2.9 m/s, CELL and SLIP technologies lowered PRA even as PLA increased, yet this reduction was large enough to lower overall predicted risk. These responses are consistent with their intended mechanisms. SLIDE and SLIP technologies incorporate sliding membrane layers and a low-friction suspended slip layer, respectively, to permit relative motion during oblique impacts [26,51]. The CELL approach uses a cellular structure that compresses and folds in shear, though its larger effect on PRA than PLA suggests the structure may have been too stiff to effectively compress under the youth impact conditions explored [22]. The rotational benefits seen across these technologies align with most adult studies of WaveCel and MIPS under oblique impacts, with the exception of Baker et al., who tested under a higher speed and different impact locations, and found stronger linear than rotational protection for WaveCel [22–24,49,52]. Jung et al. also found no overall effect of rotational technology on PRA in youth cycling helmets tested at 25°, including the same WaveCel model evaluated here [31]. The effectiveness of these commercial technologies may therefore depend strongly on impact geometry and loading conditions. By contrast, HYDRAULIC technology reduced both PLA and PRA relative to the FOAM helmets, a dual response also reported for wearable hydraulic systems in American football helmet applications [12,27,28]. This may reflect complementary deformation modes of the shock absorbers. During normal loading, compression drives controlled fluid flow that allows the shock absorber to use its available stroke, whereas during oblique loading, the fluid and surrounding compliant components can shear and permit relative motion between the shell and liner, though more work is needed to quantify this response.

The headform kinematics measured here aligned with prior bicycle helmet evaluations, with Jung et al. offering the closest comparison from tests of youth helmets on a 25° anvil at 3.1 and 5.2 m/s [31]. At 2.9 m/s, our PLA and PRA averaged 111.1 ± 27.1 g and 2756 ± 1064 rad/s$^2$, similar to the 95.9 ± 26.1 g and 3150 ± 1275 rad/s$^2$ reported at 3.1 m/s. At 5.2 m/s, PRA was nearly identical between studies, averaging 5128 ± 1888 and 4990 ± 1977 rad/s$^2$, while our PLA was lower, 125.7 ± 23.1 g versus 170.1 ± 43.5 g. Because the present study had a larger normal velocity component at 5.2 m/s, the lower PLA may reflect differences in the helmets tested and impact locations, rather than anvil angle alone. The present kinematics also fall between those reported by Hoshizaki et al. for a youth bicycle helmet, with PLA ranging from 48.1–166.0 g and PRA from 967–8158 rad/s$^2$ across front and side impacts spanning 2.0–6.0 m/s [30]. Adult bicycle studies using 6.0 to 6.5 m/s impacts against a 45° anvil have typically reported PLA ranging from 74–213.3 g and PRA from 1600–18,100 rad/s$^2$ [37,52,53]. The kinematics measured here thereby fall within the ranges observed across the broader oblique bicycle helmet literature, supporting the relevance of the selected conditions for evaluating youth helmet performance.

Greater headform displacement toward the anvil was associated with lower PLA, supporting our hypothesis that greater helmet deformation contributes to lower linear kinematics. This is mechanically plausible, since increasing the distance over which the helmet absorbs energy can reduce transmitted force, provided the liner does not densify or bottom out [48,54]. Because this displacement reflects liner compression and shell deflection, it approximates total anvil-normal deformation rather than liner compression alone. Prior work supports this mechanism, as Hansen et al. and Abayazid et al. showed that greater utilization of available liner deformation was associated with lower linear head acceleration in cycling helmets, while Marois et al. found that greater football pad deformation generally coincided with lower PLA and injury metrics [25,54,55]. Greater liner thickness has likewise been associated with lower linear acceleration in adult and youth cycling helmets, supporting greater utilization of helmet deformation as a mechanism for reducing linear head loading [31,56]. The limited deformation, and consequently higher linear accelerations, of several commercial helmets in the present study may partly reflect the impact conditions emphasized by safety certification testing, which focus on higher normal impact velocities and more concentrated loading from hemispherical and kerbstone anvils. Helmets designed to perform under these higher-severity certification impacts may therefore be relatively stiff and deform less under the lower-severity impacts evaluated here. At the same time, helmet manufacturers should ensure that any design changes leading to reduced concussion risk by increased liner deformation do not inadvertently increase the risk of superficial injury caused by contact between the helmet shell and face during liner compression, though such was not observed for any of the helmets tested here. Future helmet designs should promote sufficient deformation during concussion-relevant impacts while retaining enough available liner thickness to avoid bottoming out at higher energies.

Greater helmet headform relative rotation was associated with lower PRA, supporting our hypothesis that greater relative motion between the helmet and headform contributes to lower rotational head kinematics. Rotational decoupling is a hallmark of many advanced helmet technologies because it allows tangential deformation or sliding that limits the rotational motion transmitted to the head. Here, relative rotation averaged 5.88 ± 3.01° at 2.9 m/s and 12.52 ± 5.22° at 5.2 m/s. The 5.2 m/s measurements overlap with those reported by Zouzias et al., who measured approximately 15–21° of helmet headform relative rotation during 6.0 m/s bicycle helmet impacts onto a 45° anvil [57]. Their study also found that when a biofidelic scalp layer already permitted substantial sliding, dedicated rotational technologies added little relative motion or acceleration reduction, indicating that effectiveness depends on decoupling permitted by the complete helmet-head system. Unlike the impact mitigation systems evaluated by Zouzias et al., the HYDRAULIC system sits between the EPS liner and outer shell, providing an internal pathway for relative motion that does not require sliding at the scalp-liner interface, although overall helmet-headform motion may still depend on friction at this boundary. High speed stereo video has similarly quantified helmet head relative motion in American football impacts, with Joodaki et al. reporting much larger relative rotations up to 37° [58]. The present study extends this work by directly associating the magnitude of helmet headform relative rotation with rotational attenuation across multiple technologies and impact conditions. Future work should determine the optimal range of relative motion, since greater decoupling may improve rotational attenuation only until excessive motion compromises helmet stability, coverage, or fit.

Beyond their association with head kinematics, beta-regression analysis showed that greater helmet deformation and greater helmet-headform relative rotation were independently associated with lower predicted concussion risk, indicating that both behaviors contribute to safe helmet design. To our knowledge, this is the first cycling helmet study to directly relate both helmet deformation and relative rotation to predicted concussion risk across multiple technologies and impact conditions. Prior work has linked video- and simulation-based deformation to performance mainly in football helmets. Marois et al. found that greater pad deformation generally corresponded with lower HARM, a head impact severity metric used in American football helmet evaluations [55]. Young et al. reported strong associations between normal pad displacement and HARM at specific impact locations [59]. Bicycle helmet studies have likewise shown that greater liner deformation contributes to reduced head kinematics and brain injury metrics, although these relationships have not been directly quantified against concussion risk across helmet designs [25,54]. These findings suggest that controlled normal deformation and rotational decoupling represent complementary design objectives, and helmets that control both may provide more robust protection across diverse impacts.

The present findings have implications for both youth bicycle helmet design and evaluation. Current safety certification standards, including CPSC 1203 in the United States, evaluate impact attenuation using linear acceleration but do not assess rotational kinematics or concussion risk [9]. Under CPSC 1203, helmets pass certification when peak linear acceleration does not exceed 300 g, a criterion developed primarily to address catastrophic head injuries such as skull fractures rather than concussion. A helmet can therefore meet certification requirements while still permitting substantial concussion risk. In the present study, predicted concussion risk differed greatly across technologies, impact locations, and severity conditions, showing that these factors are not captured by a single linear-acceleration threshold. Certification standards should therefore incorporate lower impact speeds and oblique impacts generating both linear and rotational kinematics to better evaluate concussion-relevant protection. The European Committee for Standardization is already moving in this direction through revisions to EN 1078 that incorporate oblique impact testing and rotational criteria [60]. Youth-specific evaluation protocols may also be warranted as sufficient injury and exposure data become available, as demonstrated by concussion-focused helmet rating systems that separately account for youth impact exposure, anthropometry, and injury tolerance [61,62]. These data also support the goal of concussion-sensitive risk metrics that combine linear and rotational loading, which could further supplement existing criteria.

Several limitations should be considered when interpreting these findings. Only six helmet models were evaluated, reflecting the limited availability of advanced technologies in youth helmet sizes. Because each commercial technology was represented by a single helmet model, differences in performance reflect the complete helmet design and cannot be attributed solely to the incorporated technology, which is a shared limitation among prior studies investigating advanced technology in cycling helmets [22,23,63]. For example, the HYDRAULIC technology, which was prototyped into the FOAM-A helmet, may have performed differently if integrated into the FOAM-B helmet, which had a different shell geometry and EPS liner. This limitation applies to all commercially available technologies tested, as the CELL, SLIP, and SLIDE technologies may have performed differently if integrated into helmets with different shell and EPS foam characteristics as well. Additionally, because headforms differ in their frictional characteristics, use of only the NOCSAE headform leaves uncertain whether different scalp frictional properties would alter helmet-headform relative motion and rotational attenuation enough to affect comparisons among technologies [57,64]. The HYDRAULIC prototype was also not CPSC-certified at the time of experimentation, and thus would need to be certified before such a design could be made commercially available. It is possible that design changes would be necessary to enable the helmet to pass the CPSC 1203 standard that could alter the results observed in the prototype. The controlled laboratory conditions represent only a subset of the impact velocities, angles, locations, and surfaces encountered in real-world bicycle crashes. The experimental setup also excluded the neck and body mass, which prior studies have shown can have an effect on head kinematics [65–69]. Predicted concussion risk was calculated using YGAMBIT, which was developed from youth football impacts and has not been validated specifically for bicycle crashes [62]. Finally, headform displacement toward the anvil approximates helmet deformation near the impact site but does not directly measure local liner compression. Future studies should evaluate a broader range of youth helmet designs and impact conditions, directly quantify component-level deformation, and develop cycling-specific youth injury risk relationships as clinical data become available.

In summary, advanced bicycle helmet technologies significantly reduced the high predicted concussion risk experienced with conventional foam designs, although the magnitude of protection varied across technologies and impact conditions. A novel HYDRAULIC helmet provided the largest and most consistent reductions of concussion risk by attenuating both linear and rotational head kinematics across varying impact conditions. Stereo-video measurements further showed that greater headform displacement toward the anvil was associated with lower PLA and greater helmet-headform relative rotation was associated with lower PRA, identifying complementary mechanical behaviors linked to reduced concussion risk. Future youth helmet development should therefore target effective reductions in both linear and rotational loading while maintaining protection across the range of impact locations and severities encountered in cycling.

## Declarations

*Ethics approval and consent to participate*

Not Applicable

*Clinical trial number*

Not applicable

*Consent for publication*

All authors have agreed with the content of this manuscript and have given explicit consent to submit the manuscript to Annals of Biomedical Engineering.

*Availability of data and material*

Data relevant to this manuscript can be made available to interested parties upon request to the authors.

*Competing interests*

NJC and DBC have a financial interest in SoftShox and are inventors on multiple pending patent applications that claim the design of the hydraulic helmet tested in this study. FFA contributed to this study in an independent capacity, providing unpaid manuscript review and study design feedback. FFA received a one-time honorarium from SoftShox for advisory contributions, and contributed independently of any past academic or current employment affiliations. JAT was a paid consultant of SoftShox at the time of study execution and manuscript drafting. GAG has a financial interest in SoftShox. DBC's contribution to this publication was as a paid consultant and was not part of his Stanford University duties or responsibilities.

*Funding*

Funding for this study was provided in part by the National Institutes of Health under Small Business Innovation Research (SBIR) grant 2R44NS119134-03A1.

*Authors' contributions*

JAT, FFA, and NJC contributed to the study's conception and design. JAT and NJC contributed the statistical analyses in the first draft of the manuscript, with later contributions to the submitted manuscript by DBC. JAT created the first draft of the figures, with later editing by NJC. The first draft of the manuscript was written by JAT. The submitted manuscript was edited by NJC with input from DBC, GG, and FFA. All authors read and approved the final manuscript.

*Acknowledgements*

Not applicable